\documentclass{ifacconf}
\usepackage[utf8]{inputenc}

\usepackage{graphicx}      
\usepackage{natbib}        

\usepackage{amsmath} 
\usepackage{amssymb}  
\usepackage{tikz}
\usetikzlibrary{arrows, angles, quotes, external, fit, positioning, calc}
\usepackage{pgfplots}
\usepgfplotslibrary{groupplots}
\pgfplotsset{compat=1.14}
\begin{document}
\begin{frontmatter}

\title{Autonomous docking using direct optimal control} 


\author[1]{Andreas B. Martinsen} 
\author[1]{Anastasios M. Lekkas}
\author[1]{Sebastien Gros}

\address[1]{Department of Engineering Cybernetics, Norwegian University of Science and Technology (NTNU), 
   Trondheim, NO-7491 Norway (e-mail: andreas.b.martinsen@ntnu.no, anastasios.lekkas@ntnu.no, sebastien.gros@ntnu.no)}

\begin{abstract} 
We propose a method for performing autonomous docking of marine vessels using numerical optimal control. The task is framed as a dynamic positioning problem, with the addition of spatial constraints that ensure collision avoidance. The proposed method is an all-encompassing procedure for performing both docking, maneuvering, dynamic positioning and control allocation. In addition, we show that the method can be implemented as a real-time MPC-based algorithm on simulation results of a supply vessel. 
\end{abstract}


\end{frontmatter}
\section{Introduction}
For most larger vessels, docking has historically been performed by utilizing external help from support vessels such as tug boats. The main reasons for this has been limits in terms of maneuverability as well as limits in the accuracy of the human operators when dealing with relatively slow dynamical systems. With the increasing usage of azimuth thrusters, marine vessels have become increasingly maneuverable. In addition to this, interest in autonomous ferries, and cargo vessels has increased in recent years. Despite this, and contrary to topics such as path following/tracking and control allocation, research on autonomous docking for surface vessels has seen little attention. While there are some methods such as \cite{rae1992fuzzy, teo2015fuzzy, hong2003development} developed for Autonomous Underwater Vehicles (AUVs), which use fuzzy control schemes for different stages of the docking process. While \cite{breivik2011virtual} and \cite{woo2016vector} have developed methods for Unmanned Surface Vehicles (USVs) based on target tracking and artificial potential fields respectively. These existing approaches are usually quite limited, do not take into account the underlying vessel model, and make few guarantees in terms of safety.

In this paper, we present a method for framing the problem of autonomous docking as a optimal control problem. Our proposed method is similar to methods used for dynamic positioning \cite{veksler2016dynamic, sotnikova2013dynamic}, with the addition of control allocation optimization \cite{johansen2004constrained}, and spatial constraint, which ensure the vessel operates safely without colliding.
\section{Vessel Model}
\subsection{Kinematics}
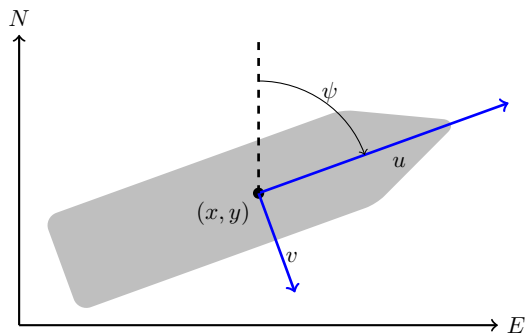
\begin{figure}
    \centering
\tikzset{
    azimuth/.pic = { 
        \draw[gray!10, fill] (-2, 3) -- (-2, -3) -- (2, -2) -- (2, 2) -- cycle;
    }
}
\tikzset{
    vessel/.pic = {
        \draw[rounded corners, gray!50, line width=2pt, fill] (39, 0) -- (20, 9) -- (-39, 9) -- (-39, -9) -- (20, -9) -- cycle;
    }
}

\tikzset{
    tunnel/.pic = {
        \draw[gray!10, line width=2pt, fill] (-1, 4) -- (1, 3) -- (1, -3) -- (-1, -4) -- cycle;
    }
}

\begin{tikzpicture}[scale = 0.07, transform shape]

    \begin{scope}[shift={(45,25)},rotate=20]
        \coordinate (body) at (0, 0);
        
        \pic at (body) {vessel};
        
        \coordinate (u) at (50, 0);
        \coordinate (v) at (0, -20);
        \coordinate (U) at ($(u) + (v)$);

        
    \end{scope}
    
    \fill (body) circle[radius=30pt];
    \node[below left, scale = 12.5] at (body) () {$(x, y)$};
    
    \draw[->, thick] (0,0)--(90,0) node[right, scale = 12.5]{$E$};
    \draw[->, thick] (0,0)--(0,55) node[above, scale = 12.5]{$N$};
    
    \coordinate (xb) at ($ (body) + (0, 30)$);
    \draw [->,line width=1pt,color=blue] (body) -- node[below right, color=black, scale = 12.5]{$u$} (u);
    \draw [->,line width=1pt,color=blue] (body) -- node[below right, color=black, scale = 12.5]{$v$} (v);
    \draw [-,line width=1pt,color=black, dashed] (body) -- node[below, color=black]{} (xb);
    
    \pic [draw, <-, "$\psi$", angle eccentricity=1.1, angle radius=1.7cm, scale = 12.5] {angle = u--body--xb};
    
\end{tikzpicture}
    \caption{3-DOF vessel centered at $(x, y)$, with surge velocity $u$, sway velocity $v$, heading $\psi$ in a North-East-Down (NED) reference frame.}
    \label{fig:3DOF_vessel}
\end{figure}

When modeling vessels for the purpose of autonomous docking, we assume the vessel moves on the ocean surface at relatively low velocities. In addition to this we assume that effects of the roll and pitch motions of the vessel are negligible, and hence have little impact on the surge, sway and yaw of the vessel. The mathematical model used to describe the system can then be kept reasonably simple by limiting it to the planar position and orientation of the vessel. The motion of a surface vessel can be represented by the pose vector $\boldsymbol{\eta} = [x, y, \psi]^\top \in \mathbb{R}^2 \times \mathbb{S}$, and velocity vector $\boldsymbol{\nu} = [u, v, r]^\top \in \mathbb{R}^3$. Here, $(x, y)$ describe the Cartesian position in the earth-fixed reference frame, $\psi$ is yaw angle, $(u, v)$ is the body fixed linear velocities, and $r$ is the yaw rate, an illustration is given in Figure \ref{fig:3DOF_vessel}.
Using the notation in \cite{fossen2011handbook} we can describe a 3-DOF vessel model as follows
\begin{align}
    &\dot{\boldsymbol{\eta}} = \boldsymbol{J}(\psi)\boldsymbol{\nu}, \\
    \boldsymbol{M}&\dot{\boldsymbol{\nu}} + \boldsymbol{D}(\boldsymbol{\nu})\boldsymbol{\nu} = \boldsymbol{\tau},
\end{align}
where $\boldsymbol{M} \in \mathbb{R}^{3 \times 3}$, $\boldsymbol{D}(\boldsymbol{\nu}) \in \mathbb{R}^{3 \times 3}$, $\boldsymbol{\tau}$ and $\boldsymbol{J}(\psi) \in SO(3)$ are the inertia matrix, dampening matrix, control input vector, and rotation matrix respectively. The rotational matrix $\boldsymbol{J}(\psi) \in SO(3)$ is given by
\begin{equation}
   \boldsymbol{J}(\psi) = 
   \begin{bmatrix}
   \cos(\psi)  & -\sin(\psi)   & 0 \\
   \sin(\psi)  & \cos(\psi)    & 0 \\
   0           & 0             & 1
   \end{bmatrix}
\end{equation}
and is the rotation from the body frame to the earth-fixed reference frame.

\subsection{Thrust configuration}
The control surfaces of the vessel are specified by the thrust configuration matrix $\boldsymbol{T}(\boldsymbol{\alpha}) \in \mathbb{R}^{3, n_{thrusters}}$ which maps the thrust $\boldsymbol{f}$ from each thruster into the surge, sway and yaw forces and moments in the body frame of the vessel given the thruster angles $\boldsymbol{\alpha}$.
\begin{equation}\label{eq:thrust_configuration}
    \boldsymbol{\tau} = \boldsymbol{T}(\boldsymbol{\alpha}) \boldsymbol{f}
\end{equation}
Each column $\boldsymbol{T}_i(\alpha_i)$ in $\boldsymbol{T}(\boldsymbol{\alpha})$ gives the configuration of the forces and moments of a thruster $i$ as follows: 
\begin{equation}
\boldsymbol{T_i}(\boldsymbol{\alpha})f_i =
    \begin{bmatrix}
    F_x \\
    F_y \\
    F_y l_x - F_x l_y
    \end{bmatrix}
    =
    \begin{bmatrix}
    f_i \cos(\alpha_i) \\
    f_i \sin(\alpha_i) \\
    f_i (l_x \sin(\alpha_i) - l_y \cos(\alpha_i))
    \end{bmatrix}
\end{equation}
where $\alpha_i$ is the orientation of the thruster in the body frame, and $f_i$ is the force it produces. Selecting the orientation $\boldsymbol{\alpha}$ and force $\boldsymbol{f}$ of the thrusters in order to generate the desired force $\boldsymbol{\tau}$ is called the thrust allocation problem. While there are numerous ways of solving the thrust allocation problem \cite{johansen2013control}, for our purpose we want to include the thrust allocation as part of the optimization for performing the docking operations. This allows us to take into account physical thruster constraints such as force saturation and feasible azimuth sectors.
\begin{equation*}
\begin{aligned}
    \alpha_{i, min} \leq \alpha_i \leq \alpha_{i, max} \\
    f_{i, min} \leq f_i \leq f_{i, max}
\end{aligned}
\end{equation*}
In order to avoid singular thruster configurations, we add a penalty on the rank deficiency of the thrust configuration matrix, as proposed by \cite{johansen2004constrained}. The singular configuration cost is given as the following.
\begin{equation}\label{eq:singular_configuration_cost}
    \frac{\rho}{\epsilon + \det \left( \boldsymbol{T}(\boldsymbol{\alpha}) \boldsymbol{W}^{-1} \boldsymbol{T}^\top (\boldsymbol{\alpha}) \right)}
\end{equation}
Here $\epsilon > 0$ is a small constant in order to avoid division by $0$, $\rho > 0$ is the weighting of the maneuverability, and $\boldsymbol{W}$ is typically diagonal matrix, weighting each individual thruster. A constraint on the singular configuration may alternatively be added, however in our implementation this is added as a cost, which means that avoiding singular thrust configurations become more important when close to the desired docking position.

It should be noted that both the singular configuration cost in (\ref{eq:singular_configuration_cost}) and the thrust configuration matrix in (\ref{eq:thrust_configuration}) are both highly nonlinear due to the trigonometric functions, adding them as costs and constraints in an optimization problem will therefor in general cause the problem to become non-convex. 
\subsection{Summary of model}
\begin{figure}
    \centering
\tikzset{
    azimuth/.pic = { 
        \draw[gray!10, fill] (-2, 3) -- (-2, -3) -- (2, -2) -- (2, 2) -- cycle;
    }
}
\tikzset{
    vessel/.pic = {
        \draw[rounded corners, gray!50, line width=2pt, fill] (39, 0) -- (20, 9) -- (-39, 9) -- (-39, -9) -- (20, -9) -- cycle;
    }
}

\tikzset{
    tunnel/.pic = {
        \draw[gray!10, line width=2pt, fill] (-1, 4) -- (1, 3) -- (1, -3) -- (-1, -4) -- cycle;
    }
}

\begin{tikzpicture}[scale = 0.1, transform shape]

    \begin{scope}[shift={(40,40)},rotate=0]
        \coordinate (body) at (0, 0);
        
        \pic at (body) {vessel};
        
        \coordinate (u) at (50, 0);
        \coordinate (v) at (0, -20);
        \coordinate (U) at ($(u) + (v)$);
        
        \coordinate (t1) at (-35, 5);
        \coordinate (t2) at (-35, -5);
        \coordinate (t3) at (30, 0);

        \pic[rotate=-30] at (t1) {azimuth};
        \node[scale = 10, right] at (t1) () {$1$}; 
        \pic[rotate=30] at (t2) {azimuth};
        \node[scale = 10, right] at (t2) () {$2$}; 
        \pic at (t3) {tunnel};
        \node[scale = 10] at (t3) () {$3$}; 
        
        \draw[<->, dashed] ([shift=(-80:4)]t1) arc (-80:260:4);
        \draw (t1) -- ++(-80:4.5) (t1) -- ++(260:4.5);
        
        \draw[<->, dashed] ([shift=(80:4)]t2) arc (80:-260:4);
        \draw (t2) -- ++(80:4.5) (t2) -- ++(-260:4.5);
        
    \end{scope}
    
    
    
    
    
\end{tikzpicture}
    \caption{Thruster configuration for vessel, where $1$ and $2$ are azimuth thrusters, and $3$ is a tunnel thruster.}
    \label{fig:thrust_configuration}
\end{figure}
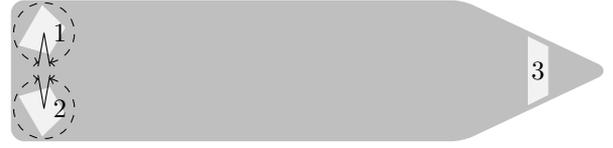

The model used for the simulations is based on the SV Northern Clipper from \cite{fossen1996identification}. The model is a 3 Degree of Freedom (3-DOF) linear model on the form:
\begin{align*}
    &\dot{\boldsymbol{\eta}} = \boldsymbol{J}(\psi)\boldsymbol{\nu}, \\
    \boldsymbol{M}&\dot{\boldsymbol{\nu}} + \boldsymbol{D}\boldsymbol{\nu} = \boldsymbol{T}(\alpha)\boldsymbol{f}
\end{align*}
For thruster configuration, we used two azimuth thrusters in the stern and one tunnel thruster in the bow, giving configuration seen in Figure \ref{fig:thrust_configuration}. Additionally saturations were added to the force generatetd by the thrusters, where the maximum thrust for the azimuth thrusters and tunnel thrusers respectively were $1/30$ and $\pm 1/60$ of the dry ships weight. For the azimuth thrusters additional constraints were added, this included a maximum turnaround time of $30s$ per revolution, and a maximum angle of $\pm 170^\circ$ giving a $20^\circ$ forbidden sector illustrated in figure \ref{fig:thrust_configuration}, which ensures the thrusters do not produce thrust that directly work against eachother, which may cause damage, this additionally reflects the movement of real world azimuth thrusters which have a finite turning radius. Additional details on the vessel model, and specific parameters are given in Appendix \ref{apx:vessel model}.

\section{Autonomous Docking}
\subsection{Obstacle avoidance}
Docking of autonomous vessels is a complex problem, which includes planning and performing maneuvers to control a vessel to a desired orientation and position, while adhering to spatial constraint in order to avoid collisions. Given a desired position $x_d, y_d$ and a desired heading $\psi_d$, we define the docking problem as maneuvering a vessel as close to the desired pose as possible, with out the vessel going aground, or running into obstacles, i.e. adhering to spatial constraints. 

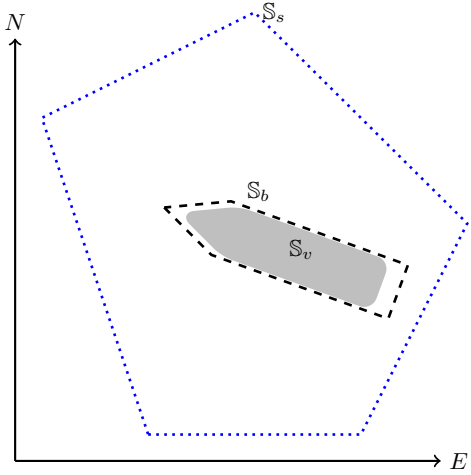
\begin{figure}
    \centering
\tikzset{
    azimuth/.pic = { 
        \draw[gray!10, fill] (-2, 3) -- (-2, -3) -- (2, -2) -- (2, 2) -- cycle;
    }
}
\tikzset{
    vessel/.pic = {
        \draw[rounded corners, gray!50, line width=2pt, fill] (39, 0) -- (20, 9) -- (-39, 9) -- (-39, -9) -- (20, -9) -- cycle;
    }
}

\tikzset{
    tunnel/.pic = {
        \draw[gray!10, line width=1pt, fill] (-1, 4) -- (1, 3) -- (1, -3) -- (-1, -4) -- cycle;
    }
}

\begin{tikzpicture}[scale = 0.07, transform shape]

    \begin{scope}[shift={(50,40)},rotate=160, scale = 0.5]
        \coordinate (body) at (0, 0);
        
        \pic at (body) {vessel};
        
        \coordinate (u) at (50, 0);
        \coordinate (v) at (0, -20);
        \coordinate (U) at ($(u) + (v)$);
        
        \draw[black, line width=1pt, scale = 1.2, dashed] (39, 0) -- (20, 9) -- (-39, 9) -- (-39, -9) -- (20, -9) -- cycle;

        
        \coordinate[scale = 1.2] (boundary) at (20, -9);
        
    \end{scope}
    
    \begin{scope}[shift={(45,5)}, scale = 0.2]
        \coordinate (spatial) at (0, 400);
        \draw[blue, line width=1pt, dotted] (-100, 0) -- (100, 0) -- (200, 200) -- (0, 400) -- (-200, 300) -- cycle;
    \end{scope}
    
    \node[above right, scale = 12.5] at (boundary) () {$\mathbb{S}_b$}; 
    \node[right, scale = 12.5] at (body) () {$\mathbb{S}_v$};
    \node[right, scale = 12.5] at (spatial) () {$\mathbb{S}_s$};
    
    \draw[->, thick] (0,0)--(80,0) node[right, scale = 12.5]{$E$};
    \draw[->, thick] (0,0)--(0,80) node[above, scale = 12.5]{$N$};

\end{tikzpicture}
    \caption{Vessel $\mathbb{S}_v$ with safety boundary $\mathbb{S}_b$ with black dashed line, and spatial constraints $\mathbb{S}_s$ as blue dotted line, in the NED frame. The vessel will always lie within the spatial constraints $\mathbb{S}_s$ as long as all the vertices of $\mathbb{S}_b$ lie within the spatial constraints.}
    \label{fig:spatial_constraints}
\end{figure}

In order to ensure the vessel does not collide we define a safety margin around the vessel which obstacles should not enter. Given a set $\mathbb{S}_v$ representing the vessel, the convex hull
\begin{equation*}
    \text{Conv}(\mathbb{S}_v)
\end{equation*}
gives the boundary points of the vessel which form a polyhedron around the vessel. By dilating the set representing the vessel by a desired safety margin $\mathbb{M}$, we get the following polyhedron representing the safety boundary surrounding the vessel.
\begin{equation}
    \mathbb{S}_b = \text{Conv}(\mathbb{S}_v \oplus \mathbb{M})
\end{equation}
For our simulations we used a safety margin of $10\%$ giving the safety boundary $\mathbb{S}_b$ seen in Figure \ref{fig:spatial_constraints}, which is a polyhedron in the body frame of the vessel consisting of five vertices.

In order to ensure safe operating conditions, we define a operating region in terms of spatial constraints $\mathbb{S}_s$ for the vessel. The operating region is chosen as the largest convex region that encompasses the desired docking position, while not intersecting with obstacles or land. Choosing the spatial constraints and vessel boundary in this way, safe operations are ensured when $\mathbb{S}_b \subseteq \mathbb{S}_s$, i.e. the vessel with the safety margin is contained within the spatial constraints, this is illustrated in Figure \ref{fig:spatial_constraints}. Using the fact that the spatial constraints are a convex polyhedron:
\begin{equation*}
    \mathbb{S}_s = \{ x | \boldsymbol{A}_s x \leq  \boldsymbol{b}_s \}
\end{equation*}
we have that the vessel is within the spatial constraints so long as all the vertices of the vessel boundary follow the linear inequality representing the spatial constraints. 
\begin{equation}
    \mathbb{S}_b \subseteq \mathbb{S}_s \Longleftrightarrow \boldsymbol{A}_s \boldsymbol{x}_i^{NED} \leq  \boldsymbol{b}_s \quad \forall \boldsymbol{x}_i^{NED} \in \text{Vertex}(\mathbb{S}_b)
\end{equation}
Since the Vertexes of the vessel boundary are given in the body frame of the vessel we need to transform them from the body frame to the NED frame, giving the following nonlinear constraints.
\begin{equation}
    \boldsymbol{A}_s \left(\boldsymbol{R}(\psi)\boldsymbol{x}_i^{b} + \begin{bmatrix} x \\ y \end{bmatrix} \right) \leq  \boldsymbol{b}_s \quad \forall \boldsymbol{x}_i^{b} \in \text{Vertex}(\mathbb{S}_b)
\end{equation}
Where $\boldsymbol{R}$ is the rotation from the body frame to NED.
\begin{equation}
   \boldsymbol{R}(\psi) = 
   \begin{bmatrix}
   \cos(\psi)  & -\sin(\psi) \\
   \sin(\psi)  & \cos(\psi) 
   \end{bmatrix}
\end{equation}
This can directly be implemented as inequality constraints in an optimization problem, and ensures the vessel is contained within a predefined safe region.

While this constraint is easily implemented in a nonlinear programming (NLP) problem, the constraint is not convex. This means the constraint will enforce safety requirements, however the NLP may not converge to a global optimum.

\subsection{Optimal control problem (OCP)}
Using the model, and constraints discussed in the previous sections, with the desired docking pose $\boldsymbol{\eta}_d = [x_d, y_d, \psi_d]^\top$, we can formulate the following nonlinear continuous time optimal control problem.
\begin{subequations}
\begin{align}
    J^* = \min_{\boldsymbol{\eta}, \boldsymbol{\nu}, \boldsymbol{f}, \boldsymbol{\alpha}} & \int_{0}^{T} \Bigg{\{} 
    ||\boldsymbol{\eta} - \boldsymbol{\eta}_d||^2_{\boldsymbol{Q}_{\eta}} + ||\boldsymbol{\nu}||^2_{\boldsymbol{Q}_{\nu}} +  ||\boldsymbol{f}||^2_{\boldsymbol{R}_{f}} + \notag \\
    & \quad \quad \frac{\rho}{\epsilon + \det \left( \boldsymbol{T}(\boldsymbol{\alpha}) \boldsymbol{W}^{-1} \boldsymbol{T}^\top (\boldsymbol{\alpha}) \right)} 
    \Bigg{\}}dt \label{eq:ocp_objective} \\
    \text{subject to:} \notag \\
    & \dot{\boldsymbol{\eta}} = \boldsymbol{J}(\psi)\boldsymbol{\nu} \label{eq:ocp_model_1}\\
    & \boldsymbol{M}\dot{\boldsymbol{\nu}} + \boldsymbol{D}\boldsymbol{\nu} = \boldsymbol{T}(\alpha)\boldsymbol{f} \label{eq:ocp_model_2}\\
    &     \boldsymbol{A}_s \left(\boldsymbol{R}(\psi)\boldsymbol{x}_i^{b} + \begin{bmatrix} x \\ y \end{bmatrix} \right) \leq  \boldsymbol{b}_s \: \forall \boldsymbol{x}_i^{b} \in \text{Vertex}(\mathbb{S}_b) \label{eq:ocp_spatial}\\
    & \boldsymbol{f}_{min} \leq \boldsymbol{f} \leq \boldsymbol{f}_{max} \label{eq:ocp_force}\\
    & \boldsymbol{\alpha}_{min} \leq \boldsymbol{\alpha} \leq \boldsymbol{\alpha}_{max} \label{eq:ocp_orientation}\\
    & |\dot{\boldsymbol{\alpha}}| \leq \dot{\boldsymbol{\alpha}}_{max} \label{eq:ocp_orientation_derivative}\\
    & \text{Initial conditions on } \boldsymbol{\eta}, \boldsymbol{\nu}, \boldsymbol{f}, \boldsymbol{\alpha} \label{eq:ocp_initial_conditions}
\end{align}
\end{subequations}
Where we minimize cost (\ref{eq:ocp_objective}), subject to the dynamic model constraints (\ref{eq:ocp_model_1}) and (\ref{eq:ocp_model_2}), the spatial constraints (\ref{eq:ocp_spatial}), the saturation constraint (\ref{eq:ocp_force}), (\ref{eq:ocp_orientation}) and (\ref{eq:ocp_orientation_derivative}), and the initial conditions (\ref{eq:ocp_initial_conditions}) over the time horizon $T$. For this problem we have opted to use a simple quadratic penalty in order to ensure the vessel converges to the desired pose, however Huber penalty functions as discussed in \cite{6580317, 7807211} may give better performance for large pose deviations.

\subsection{Implementation}
In order to implement the proposed docking system we need to solve the OCP in the previous section. This can be done in multiple ways, however the two main classes of methods are sequential methods, such as direct single shooting \cite{hicks1971approximation}, and simultaneous methods such as direct multiple shooting \cite{deuflhard1974modified}, and direct collocation \cite{tsang1975optimal}. For this approach we chose to use direct collocation, in where implicit numerical integration of the ODE constraints (\ref{eq:ocp_model_1}) and (\ref{eq:ocp_model_2}), as well as the objective function (\ref{eq:ocp_objective}), is performed as part of the nonlinear optimization. In the collocation method, the numerical integration is performed by fitting the derivatives to a degree $d$ Legendre polynomial, with known integral, within $N$ set time intervals called shooting intervals. The shooting intervals are then connected to create the full time horizon, by enforcing constraints on the shooting gaps between intervals.

For this problem we opted to use direct collocation for several reasons. Comparing direct collocation with multiple shooting, they both offer the same stability in terms of the optimization, however direct collocation offers a speedup, as the numerical integration is performed as part of the optimization, and not offloaded to a separate integration routine, giving the optimization problem a nice sparsity structure. While multiple shooting offers more flexibility in terms of the integrator used, the implicit integrator of the direct collocation is sufficient for our purpose. Comparing single shooting to direct collocation the single shooting problem has much fewer decision variables, however the problem often becomes very dense, and hence increases the computation time, single shooting is also more unstable, as propagating the gradients through a long time horizon often cause them to become very small (vanish) or very large (explode), and hence the optimization steps may be oscillatory and unstable.

For the implementation we used CasADi \cite{Andersson2018} a software framework for easy implementation of nonlinear optimization and optimal control problems, with IPOPT \cite{wachter2006implementation} an interior point optimizer, for solving the resulting NLP. 

Solving the OCP once, gives a open loop trajectory over a time horizon $T$, which can be used to perform open loop control, or trajectory tracking. We however wish to use the OCP as the basis for a Nonlinear Model Predictive Control (NLMPC). Where at each time step the OCP is solved with the vessel state as initial conditions, and then only the first predicted control action is performed. This gives a closed loop control scheme, which makes the method more robust to modeling errors, and external disturbances due to the feedback. 

\section{Simulation}
\begin{figure}
    \centering
\begin{tikzpicture}

\definecolor{color0}{rgb}{0.12156862745098,0.466666666666667,0.705882352941177}
\definecolor{color1}{rgb}{1,0.498039215686275,0.0549019607843137}

\begin{axis}[
ticks = none,
x grid style={white!69.01960784313725!black},
y grid style={white!69.01960784313725!black},
ymin=50, ymax = 800,
xmin=0, xmax = 1100,
unit vector ratio*=1 1 1,
yscale = -1,
]
\addplot[yshift=159, yscale=-1] graphics [includegraphics cmd=\pgfimage,xmin=-0.5, xmax=1699.5, ymin=995.5, ymax=-0.5] {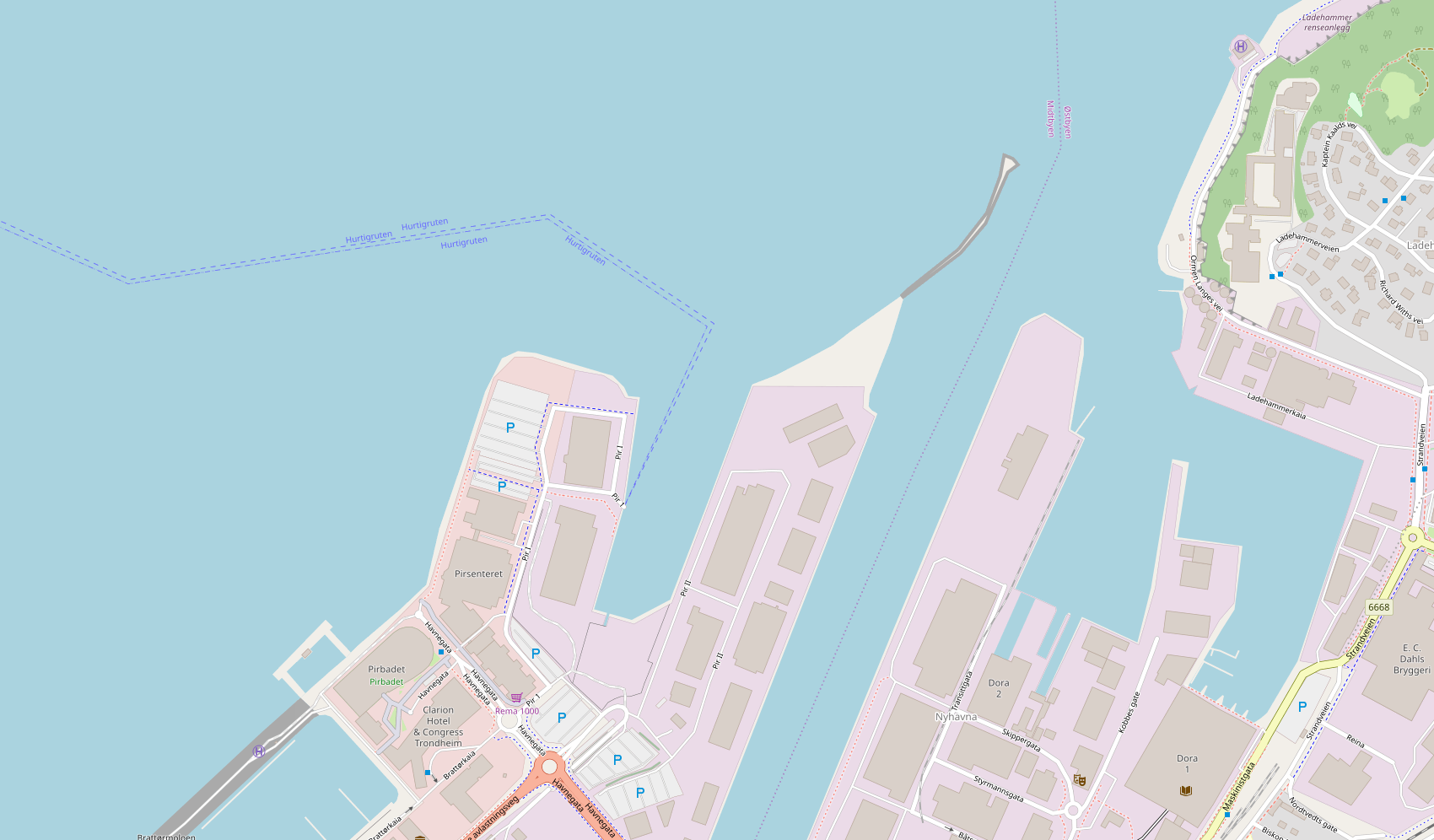};
\addplot [semithick, color0, forget plot]
table [row sep=\\]{%
900	200 \\
899.448589654661	202.837279054626 \\
897.513900478129	210.840145308494 \\
893.605397614646	223.206845953998 \\
887.600928723729	239.256890812998 \\
879.993309226628	258.631928647168 \\
871.346645509504	281.130311748992 \\
862.207647026811	306.583154592552 \\
853.057962732485	334.79199206903 \\
844.211032601006	365.430610976752 \\
835.950267550295	396.761101622084 \\
828.760119111862	426.267877887859 \\
822.804648289153	452.323621359204 \\
817.928608012182	474.106455240846 \\
813.750842037168	491.471093277501 \\
809.751874064609	504.898771466522 \\
805.537690387768	515.226247205331 \\
800.917471094939	523.216666339994 \\
795.894957164615	529.374264253099 \\
790.67584994944	534.033322566982 \\
785.561803542055	537.498464317156 \\
780.775199382879	540.064082697595 \\
776.48932496485	542.028131763294 \\
772.782102320407	543.61141429677 \\
769.66414995788	544.950195357858 \\
767.09103376881	546.115605605977 \\
764.990834139631	547.144221431849 \\
763.302692854268	548.066120859024 \\
761.960037510065	548.8642745597 \\
760.896578435283	549.522845119427 \\
760.043943169802	550.045101104629 \\
759.349799734886	550.444806237841 \\
758.782471986156	550.73901650295 \\
758.321480231536	550.946291500801 \\
757.950221572093	551.084414525772 \\
757.653478022183	551.169095795125 \\
757.416807660081	551.21247051162 \\
757.229435321788	551.225408297124 \\
757.081739823304	551.217378739325 \\
756.965121567339	551.196592320541 \\
756.87361894259	551.169084782869 \\
756.802422794552	551.138299267822 \\
756.747376969856	551.106789762989 \\
756.704949668829	551.076444265101 \\
756.672447937346	551.048566221156 \\
756.648076503915	551.02382468728 \\
756.631567530439	551.002358860982 \\
756.621206662137	550.984016074116 \\
756.614333060074	550.968802684801 \\
756.610101238428	550.956439656142 \\
756.607890796125	550.946693312415 \\
756.607064568712	550.939669726264 \\
756.607167000386	550.934945138726 \\
756.607571306431	550.931706835622 \\
756.640781042005	550.941361434935 \\
756.67934095055	550.955596826778 \\
756.685054132981	550.960370877777 \\
756.679134926016	550.961205440469 \\
756.672464522957	550.962591042759 \\
757.390730317878	550.880182616367 \\
758.465735961016	550.726394081856 \\
};
\addplot [semithick, color1, forget plot]
table [row sep=\\]{%
748	734 \\
793	710 \\
1030	100 \\
810	100 \\
722	726 \\
748	734 \\
};
\addplot [thin, black, forget plot, fill=black, fill opacity=0.2]
table [row sep=\\]{%
900	242.9 \\
890.1	222 \\
890.1	157.1 \\
909.9	157.1 \\
909.9	222 \\
900	242.9 \\
};
\addplot [thin, black, forget plot, fill=black, fill opacity=0.2]
table [row sep=\\]{%
819.427921267301	436.351783269472 \\
818.340958307055	413.251166145182 \\
843.336302683891	353.357570832466 \\
861.608924982685	360.983269116925 \\
836.613580605849	420.87686442964 \\
819.427921267301	436.351783269472 \\
};
\addplot [thin, black, forget plot, fill=black, fill opacity=0.2]
table [row sep=\\]{%
753.108247585035	509.441794775947 \\
775.393570381298	515.62114651107 \\
824.489975547045	558.065851714437 \\
811.540743451103	573.044416002293 \\
762.444338285355	530.599710798925 \\
753.108247585035	509.441794775947 \\
};
\addplot [thin, black, forget plot, fill=black, fill opacity=0.2]
table [row sep=\\]{%
759.382028337471	507.150207833346 \\
769.603321190184	527.894970363176 \\
770.604679526275	592.787244799219 \\
750.807036477991	593.092743952603 \\
749.8056781419	528.20046951656 \\
759.382028337471	507.150207833346 \\
};
\addplot [thin, black, forget plot, fill=black, fill opacity=0.2]
table [row sep=\\]{%
762.270743146315	508.60993711366 \\
769.462716714318	530.589371050801 \\
761.297836508681	594.973722652938 \\
741.655152969046	592.482742251219 \\
749.820033174683	528.098390649082 \\
762.270743146315	508.60993711366 \\
};
\addplot [thin, black, forget plot, fill=black, fill opacity=0.2]
table [row sep=\\]{%
762.575210819647	508.463742711769 \\
769.47181274167	530.537638650845 \\
760.44432860352	594.806717764642 \\
740.836812941684	592.052570061478 \\
749.864297079833	527.783490947681 \\
762.575210819647	508.463742711769 \\
};




\end{axis}

\end{tikzpicture}
    \caption{Vessel docking performed at Hurtigruten terminal in Trondheim Norway.}
    \label{fig:docking_trondheim}
\end{figure}
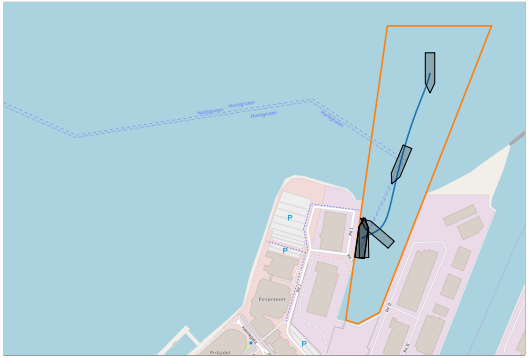

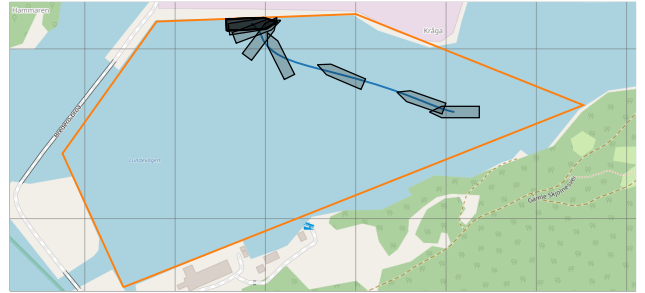
\begin{figure}
    \centering
\begin{tikzpicture}[scale=1.2]

\definecolor{color0}{rgb}{0.12156862745098,0.466666666666667,0.705882352941177}
\definecolor{color1}{rgb}{1,0.498039215686275,0.0549019607843137}

\begin{axis}[
ticks=none,
x grid style={white!69.01960784313725!black},
y grid style={white!69.01960784313725!black},
ymin=360, ymax = 860,
unit vector ratio*=1 1 1,
yscale = -1,
]

\addplot[yshift=42, yscale=-1] graphics [includegraphics cmd=\pgfimage,xmin=-0.5, xmax=1664.5, ymin=950.5, ymax=-0.5] {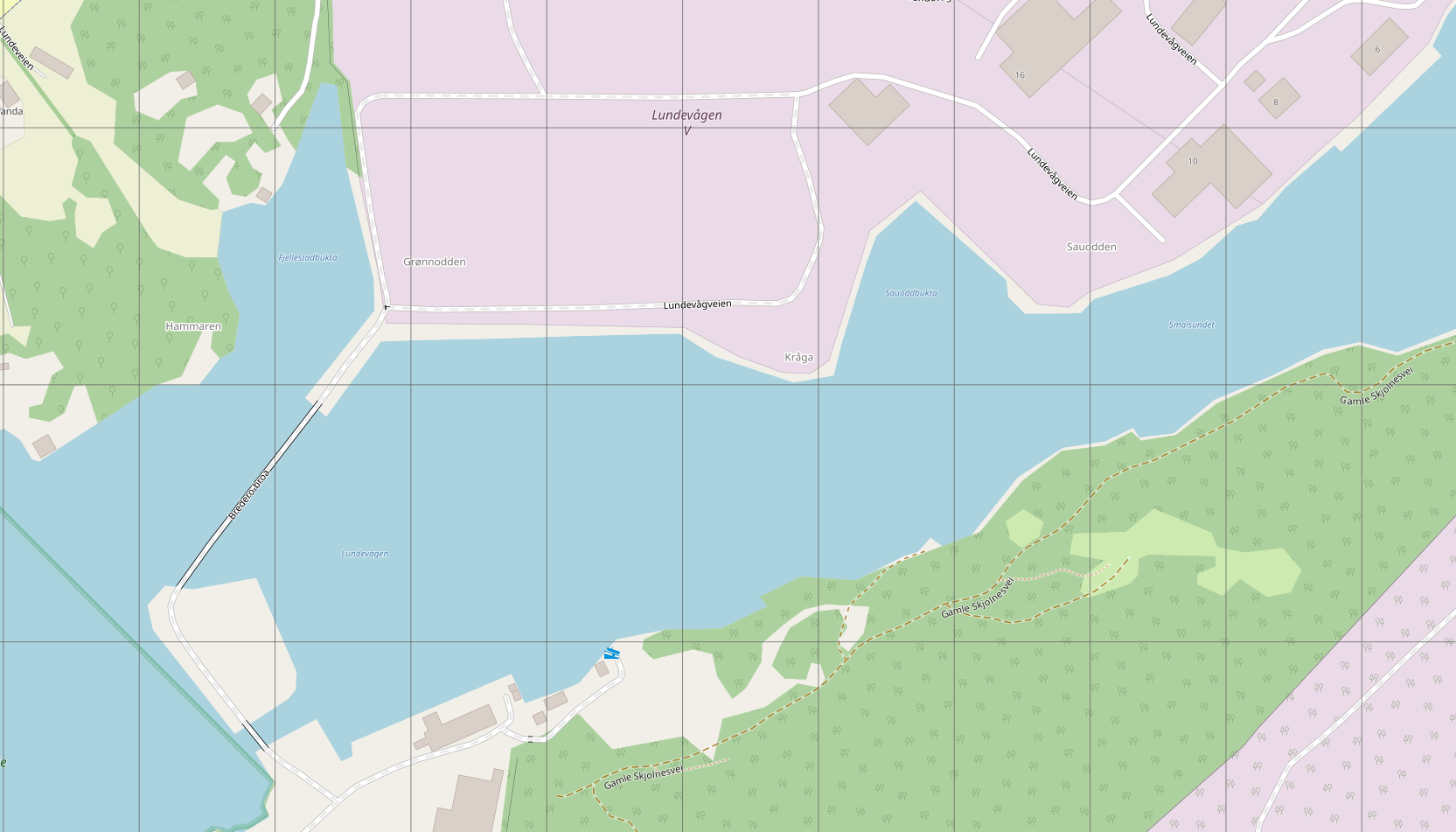};

\addplot [semithick, color0, forget plot]
table [row sep=\\]{%
950	550 \\
947.162518271327	549.444752967938 \\
939.158435857708	547.505943831655 \\
926.794888659145	543.566535874913 \\
910.768509075406	537.471915182834 \\
891.437533129819	529.73224186044 \\
868.993827853191	520.929603230177 \\
843.596274664401	511.6256138697 \\
815.438093218113	502.309859883278 \\
784.853429854405	493.280609268835 \\
753.621346485673	484.754374819951 \\
724.307004803377	477.085751812146 \\
698.547709581287	470.234096056117 \\
677.08383940533	463.798311439452 \\
659.979650145928	457.34427234619 \\
646.823307858602	450.697826024959 \\
636.968712847528	444.012529272685 \\
629.697734887678	437.449993346819 \\
624.361476326973	431.160694976762 \\
620.740876506711	425.278635494042 \\
618.532279019917	420.058723021359 \\
616.913091428603	415.695862752058 \\
615.435067333986	412.542833514557 \\
613.959688777231	410.205342956137 \\
612.421267708601	408.066665567729 \\
610.851342810063	406.1518652981 \\
609.298297202494	404.462860283388 \\
607.80972358612	402.991752080514 \\
606.42475079718	401.726296351388 \\
605.17146093775	400.652020154523 \\
604.066854704366	399.753033845855 \\
603.117539064414	399.01257318958 \\
602.321860132582	398.41375087304 \\
601.670048905602	397.939809138717 \\
601.148250259238	397.573302957314 \\
600.741342986199	397.297156161804 \\
600.432023360675	397.092980710239 \\
600.196811684354	396.947585534947 \\
600.016480768469	396.850917213936 \\
599.888563275935	396.790272840681 \\
599.809245997257	396.756186910597 \\
599.767387188922	396.741518374615 \\
599.753304263087	396.738610053213 \\
599.769352485916	396.744383159384 \\
599.804660842697	396.754056149357 \\
599.841107854471	396.763056104904 \\
599.910320456024	396.788385719819 \\
599.999029837949	396.82285512422 \\
600.063351751954	396.843823655705 \\
600.105392521081	396.853275271457 \\
600.129998569342	396.856566483701 \\
600.141416226467	396.857156377892 \\
600.143218687624	396.856484571446 \\
600.138349885637	396.854885152332 \\
600.129180967307	396.852721474006 \\
600.115753998471	396.839563960411 \\
600.100137771159	396.819991491764 \\
600.085689723221	396.806679006141 \\
600.072922990087	396.799098315523 \\
600.061938446425	396.795881686833 \\
600.052758936833	396.795369339861 \\
};
\addplot [semithick, color1, forget plot]
table [row sep=\\]{%
380	854 \\
275	622 \\
437	393 \\
780	380 \\
1172	538 \\
380	854 \\
};
\addplot [thin, black, forget plot, fill=black, fill opacity=0.2]
table [row sep=\\]{%
907.1	550 \\
928	540.1 \\
992.9	540.1 \\
992.9	559.9 \\
928	559.9 \\
907.1	550 \\
};
\addplot [thin, black, forget plot, fill=black, fill opacity=0.2]
table [row sep=\\]{%
850.963841565143	515.509362717051 \\
873.963996745126	513.098375015571 \\
935.193427573739	534.615038335057 \\
928.629021815251	553.2952036726 \\
867.399590986639	531.778540353114 \\
850.963841565143	515.509362717051 \\
};
\addplot [thin, black, forget plot, fill=black, fill opacity=0.2]
table [row sep=\\]{%
714.189044647295	467.857523878812 \\
737.298926529332	466.989561092562 \\
796.95292161816	492.551463798387 \\
789.154375029942	510.750987723792 \\
729.500379941114	485.189085017967 \\
714.189044647295	467.857523878812 \\
};
\addplot [thin, black, forget plot, fill=black, fill opacity=0.2]
table [row sep=\\]{%
627.670086521093	412.310795501664 \\
645.859688575511	426.592195191536 \\
674.835074701486	484.664882393444 \\
657.117983690735	493.504830703064 \\
628.14259756476	435.432143501156 \\
627.670086521093	412.310795501664 \\
};
\addplot [thin, black, forget plot, fill=black, fill opacity=0.2]
table [row sep=\\]{%
644.32786654918	385.780484093139 \\
639.671148274769	408.432966846973 \\
600.647054320244	460.289789840947 \\
584.826328661065	448.384134058211 \\
623.850422615591	396.527311064236 \\
644.32786654918	385.780484093139 \\
};
\addplot [thin, black, forget plot, fill=black, fill opacity=0.2]
table [row sep=\\]{%
650.968999994728	390.953721266223 \\
634.931764348017	407.615866170522 \\
574.240949632754	430.607930218745 \\
567.226421618042	412.092088441207 \\
627.917236333304	389.100024392984 \\
650.968999994728	390.953721266223 \\
};
\addplot [thin, black, forget plot, fill=black, fill opacity=0.2]
table [row sep=\\]{%
646.451419932679	393.122903348885 \\
627.332559295109	406.134020438558 \\
563.212319590738	416.16421785705 \\
560.152259361367	396.602110828598 \\
624.272499065739	386.571913410106 \\
646.451419932679	393.122903348885 \\
};
\addplot [thin, black, forget plot, fill=black, fill opacity=0.2]
table [row sep=\\]{%
643.550634008192	394.508858797056 \\
623.338278799599	405.746327236239 \\
558.575505202224	409.964520685474 \\
557.288598726187	390.206386367631 \\
622.051372323561	385.988192918396 \\
643.550634008192	394.508858797056 \\
};
\addplot [thin, black, forget plot, fill=black, fill opacity=0.2]
table [row sep=\\]{%
642.677019043365	395.093648469734 \\
622.176382071358	405.79616610541 \\
557.325135668271	408.311288362099 \\
556.557810234027	388.526162340819 \\
621.409056637114	386.011040084129 \\
642.677019043365	395.093648469734 \\
};
\addplot [thin, black, forget plot, fill=black, fill opacity=0.2]
table [row sep=\\]{%
642.711371054048	395.166017220286 \\
622.194405648141	405.837199466541 \\
557.339392089806	408.253232650963 \\
556.602297219982	388.466957328082 \\
621.457310778317	386.050924143659 \\
642.711371054048	395.166017220286 \\
};
\addplot [thin, black, forget plot, fill=black, fill opacity=0.2]
table [row sep=\\]{%
643.003520612759	395.349547043673 \\
622.464193846998	405.977625703962 \\
557.604250242854	408.257475626057 \\
556.908702808995	388.469696221403 \\
621.768646413139	386.189846299308 \\
643.003520612759	395.349547043673 \\
};
\addplot [thin, black, forget plot, fill=black, fill opacity=0.2]
table [row sep=\\]{%
642.988513441776	395.311003292173 \\
622.45452925181	405.949400412334 \\
557.595739324759	408.261838346335 \\
556.890249785572	388.474410910963 \\
621.749039712623	386.161972976962 \\
642.988513441776	395.311003292173 \\
};




\end{axis}

\end{tikzpicture}
    \caption{Vessel docking performed at Lundevågen harbour in Farsund Norway.}
    \label{fig:docking_lundevaagen}
\end{figure}

\begin{figure}
    \centering
    \input{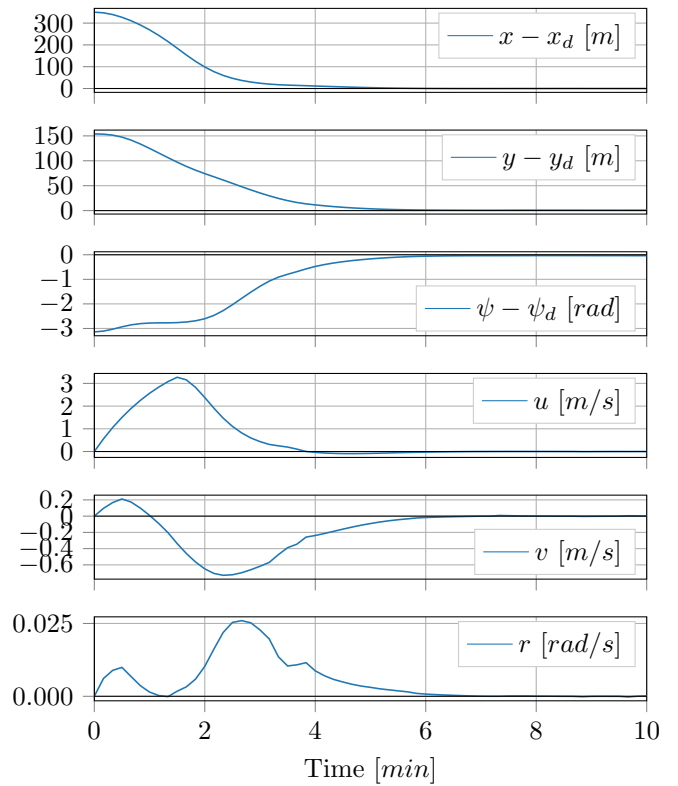}
    \caption{Vessel pose error $\boldsymbol{\eta} - \boldsymbol{\eta}_d$ and velocity $\boldsymbol{\nu}$ when docking at Lundevågen harbour.}
    \label{fig:lundevaagen_state}
\end{figure}

\begin{figure}
    \centering
    \input{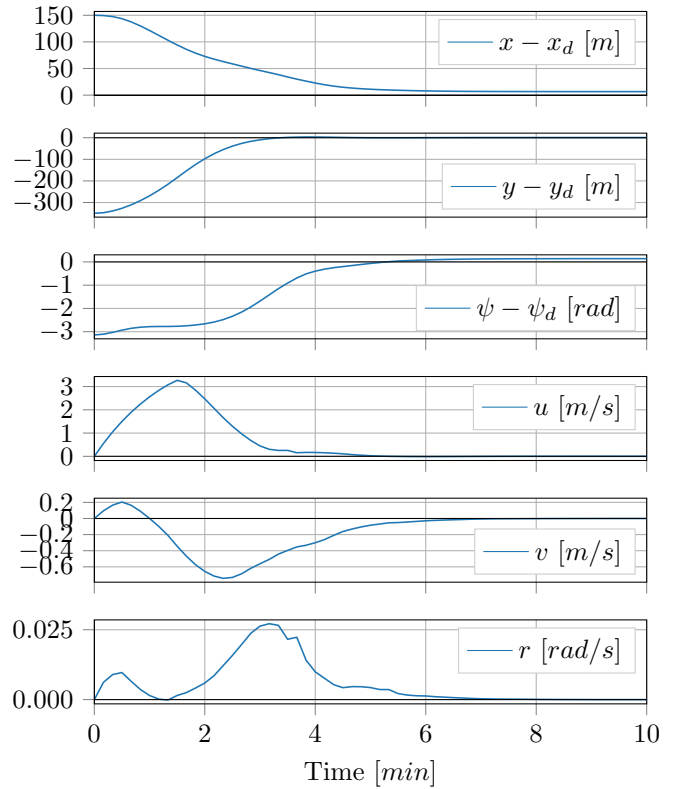}
    \caption{Vessel pose error $\boldsymbol{\eta} - \boldsymbol{\eta}_d$ and velocity $\boldsymbol{\nu}$ when docking at Trondheim harbour.}
    \label{fig:trondheim_state}
\end{figure}

\begin{figure}
    \centering
    \input{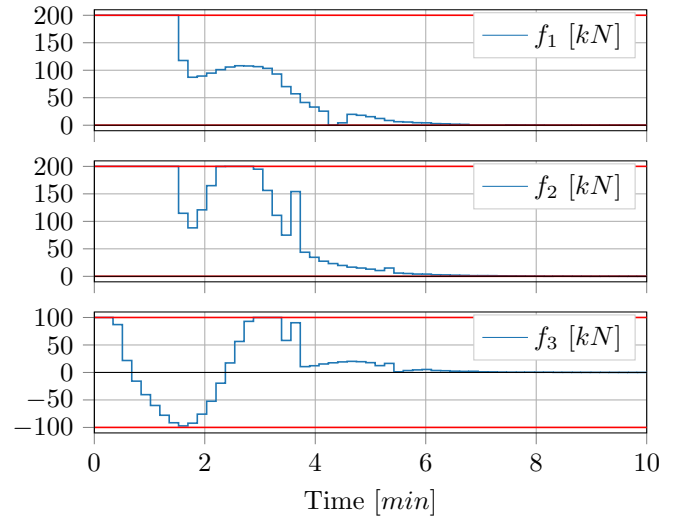}
    \caption{Thruster force for docking at Trondheim harbour, with saturation constraints indicated in red.}
    \label{fig:trondheim_force}
\end{figure}

\begin{figure}
    \centering
    \input{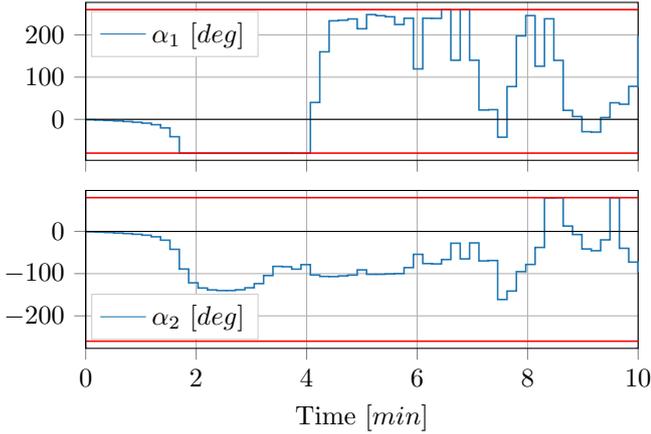}
    \caption{Azimuth angles when docking at Trondheim harbour, with saturation constraints indicated in red.}
    \label{fig:trondheim_azimuth}
\end{figure}

As a proof of concept, simulations were performed, where the OCP was run as a closed loop Nonlinear Model Predictive Control (NLMPC). For the OCP we used a time horizon of $T = 300$ seconds, with $N = 30$ time steps, making each time step $T/N = 10$ seconds. Using this we performed docking simulations at two different locations, namely Trondheim harbour and Lundevågen harbour, as seen in Figure \ref{fig:docking_trondheim} and \ref{fig:docking_lundevaagen} respectively. For the docking at Lundevågen harbour, the vessel state and control inputs are shown in Figure \ref{fig:lundevaagen_state}, \ref{fig:lundevaagen_force} and \ref{fig:lundevaagen_azimuth}, and for the docking at Trondheim harbour, the vessel state and control inputs are shown in Figure \ref{fig:trondheim_state}, \ref{fig:trondheim_force} and \ref{fig:trondheim_azimuth}. From the simulations we see an expected behaviour, where the vessel will turn and face the bow in the direction of travel, as this is the most efficient way of traveling. As the vessel closes in on the target position, it will start initiating the turn such that it faces in the desired heading, while simultaneously adhering to the defined spatial constraints in order to avoid colliding.

\begin{figure}
    \centering
    \input{figures/lundevaagen_force.tex}
    \caption{Thruster force for docking at Lundevågen harbour, with saturation constraints indicated in red.}
    \label{fig:lundevaagen_force}
\end{figure}

\begin{figure}
    \centering
    \input{figures/lundevaagen_azimuth.tex}
    \caption{Azimuth angles when docking at Lundevågen harbour, with saturation constraints indicated in red.}
    \label{fig:lundevaagen_azimuth}
\end{figure}

\section{Conclusion}
Based on the results of the simulation, the proposed method works very well, with the vessel approaching the target poses without violating the spatial constraints. Solving the open loop optimization problem with zeros as a trivial initial guess takes $2 - 4$ seconds, while solving the problem using a warm start, a solution is found in about $0.5$ seconds. With a purpose build solver this should take even less time, and ensures real time feasibility, as demonstrated by \cite{VUKOV201564}. NLP solver for the problem should be chosen carefully. We fond that IPOPT worked the best, as it was able to consistently solve the problem from a number of tested initial points, within a reasonable amount of time. With other solvers outright failing, or using excessive amounts of time.

The method does however have some drawbacks, since the proposed problem is non-convex due to the rotation of the azimuth thrusters and the vessel rotation, this means convergence to a global optimum can not be guaranteed. The method will however converge to a locally optimal solution, which in practise may be good enough, and will most importantly ensure safe operations. It is also worth noting that the problem is a finite horizon optimization problem, meaning we are only optimizing over a horizon $T$. This means that maneuvers that are optimal over a time horizon longer than $T$, may no longer be optimal over $T$, meaning the horizon must also be carefully chosen to get the desired behaviour.

The proposed method seems very promising, however many improvements can be made. Future research can be done on using more complex nonlinear vessel models, which may include thrust and azimuth dynamics. Different objective functions may be implemented, such as minimizing time, until the vessel reaches a terminal set, or energy expended. The method may also be further generalized by having dynamic spatial constraints, that use the largest convex set that does not intersect obstacles centered about the vessel as constraints. This may make the method not only suitable for docking, but also for general obstacle avoidance while in transit. While the proposed docking method has some measures ensuring robustness and safety while performing docking, future research can be done into making the method able to handle external environmental forces such as wind waves and currents using for example a scenario based MPC.

\bibliography{ifacconf}            

\begin{thebibliography}{20}
\providecommand{\natexlab}[1]{#1}
\providecommand{\url}[1]{\texttt{#1}}
\providecommand{\urlprefix}{URL }
\expandafter\ifx\csname urlstyle\endcsname\relax
  \providecommand{\doi}[1]{doi:\discretionary{}{}{}#1}\else
  \providecommand{\doi}{doi:\discretionary{}{}{}\begingroup
  \urlstyle{rm}\Url}\fi

\bibitem[{Andersson et~al.(In Press, 2018)Andersson, Gillis, Horn, Rawlings,
  and Diehl}]{Andersson2018}
Andersson, J.A.E., Gillis, J., Horn, G., Rawlings, J.B., and Diehl, M. (In
  Press, 2018).
\newblock {CasADi} -- {A} software framework for nonlinear optimization and
  optimal control.
\newblock \emph{Mathematical Programming Computation}.

\bibitem[{Breivik and Loberg(2011)}]{breivik2011virtual}
Breivik, M. and Loberg, J.E. (2011).
\newblock A virtual target-based underway docking procedure for unmanned
  surface vehicles.
\newblock \emph{IFAC Proceedings Volumes}, 44(1), 13630--13635.

\bibitem[{Deuflhard(1974)}]{deuflhard1974modified}
Deuflhard, P. (1974).
\newblock A modified newton method for the solution of ill-conditioned systems
  of nonlinear equations with application to multiple shooting.
\newblock \emph{Numerische Mathematik}, 22(4), 289--315.

\bibitem[{Fossen and Perez(2004)}]{fossenperez2004}
Fossen, T.I. and Perez, T. (2004).
\newblock Marine systems simulator (mss).
\newblock \urlprefix\url{http://www.marinecontrol.org}.

\bibitem[{Fossen(2011)}]{fossen2011handbook}
Fossen, T.I. (2011).
\newblock \emph{Handbook of marine craft hydrodynamics and motion control}.
\newblock John Wiley \& Sons.

\bibitem[{Fossen et~al.(1996)Fossen, Sagatun, and
  S{\o}rensen}]{fossen1996identification}
Fossen, T.I., Sagatun, S.I., and S{\o}rensen, A.J. (1996).
\newblock Identification of dynamically positioned ships.

\bibitem[{Gros and Diehl(2013)}]{6580317}
Gros, S. and Diehl, M. (2013).
\newblock Nmpc based on huber penalty functions to handle large deviations of
  quadrature states.
\newblock In \emph{2013 American Control Conference}, 3159--3164.
\newblock \doi{10.1109/ACC.2013.6580317}.

\bibitem[{Gros and Zanon(2017)}]{7807211}
Gros, S. and Zanon, M. (2017).
\newblock Penalty functions for handling large deviation of quadrature states
  in nmpc.
\newblock \emph{IEEE Transactions on Automatic Control}, 62(8), 3848--3860.
\newblock \doi{10.1109/TAC.2017.2649043}.

\bibitem[{Hicks and Ray(1971)}]{hicks1971approximation}
Hicks, G. and Ray, W. (1971).
\newblock Approximation methods for optimal control synthesis.
\newblock \emph{The Canadian Journal of Chemical Engineering}, 49(4), 522--528.

\bibitem[{Hong et~al.(2003)Hong, Kim, Oh, Lee, Jeon, Oh
  et~al.}]{hong2003development}
Hong, Y.H., Kim, J.Y., Oh, J.h., Lee, P.M., Jeon, B.H., Oh, K.H., et~al.
  (2003).
\newblock Development of the homing and docking algorithm for auv.
\newblock In \emph{The Thirteenth International Offshore and Polar Engineering
  Conference}. International Society of Offshore and Polar Engineers.

\bibitem[{Johansen and Fossen(2013)}]{johansen2013control}
Johansen, T.A. and Fossen, T.I. (2013).
\newblock Control allocation—a survey.
\newblock \emph{Automatica}, 49(5), 1087--1103.

\bibitem[{Johansen et~al.(2004)Johansen, Fossen, and
  Berge}]{johansen2004constrained}
Johansen, T.A., Fossen, T.I., and Berge, S.P. (2004).
\newblock Constrained nonlinear control allocation with singularity avoidance
  using sequential quadratic programming.
\newblock \emph{IEEE Transactions on Control Systems Technology}, 12(1),
  211--216.

\bibitem[{Rae and Smith(1992)}]{rae1992fuzzy}
Rae, G. and Smith, S. (1992).
\newblock A fuzzy rule based docking procedure for autonomous underwater
  vehicles.
\newblock In \emph{OCEANS 92 Proceedings@ m\_Mastering the Oceans Through
  Technology}, volume~2, 539--546. IEEE.

\bibitem[{Sotnikova and Veremey(2013)}]{sotnikova2013dynamic}
Sotnikova, M.V. and Veremey, E.I. (2013).
\newblock Dynamic positioning based on nonlinear mpc.
\newblock \emph{IFAC Proceedings Volumes}, 46(33), 37--42.

\bibitem[{Teo et~al.(2015)Teo, Goh, and Chai}]{teo2015fuzzy}
Teo, K., Goh, B., and Chai, O.K. (2015).
\newblock Fuzzy docking guidance using augmented navigation system on an auv.
\newblock \emph{IEEE journal of oceanic engineering}, 40(2), 349--361.

\bibitem[{Tsang et~al.(1975)Tsang, Himmelblau, and Edgar}]{tsang1975optimal}
Tsang, T., Himmelblau, D., and Edgar, T. (1975).
\newblock Optimal control via collocation and non-linear programming.
\newblock \emph{International Journal of Control}, 21(5), 763--768.

\bibitem[{Veksler et~al.(2016)Veksler, Johansen, Borrelli, and
  Realfsen}]{veksler2016dynamic}
Veksler, A., Johansen, T.A., Borrelli, F., and Realfsen, B. (2016).
\newblock Dynamic positioning with model predictive control.
\newblock \emph{IEEE Transactions on Control Systems Technology}, 24(4),
  1340--1353.

\bibitem[{Vukov et~al.(2015)Vukov, Gros, Horn, Frison, Geebelen, Jørgensen,
  Swevers, and Diehl}]{VUKOV201564}
Vukov, M., Gros, S., Horn, G., Frison, G., Geebelen, K., Jørgensen, J.,
  Swevers, J., and Diehl, M. (2015).
\newblock Real-time nonlinear mpc and mhe for a large-scale mechatronic
  application.
\newblock \emph{Control Engineering Practice}, 45, 64 -- 78.
\newblock \doi{https://doi.org/10.1016/j.conengprac.2015.08.012}.
\newblock
  \urlprefix\url{http://www.sciencedirect.com/science/article/pii/S0967066115300095}.

\bibitem[{W{\"a}chter and Biegler(2006)}]{wachter2006implementation}
W{\"a}chter, A. and Biegler, L.T. (2006).
\newblock On the implementation of an interior-point filter line-search
  algorithm for large-scale nonlinear programming.
\newblock \emph{Mathematical programming}, 106(1), 25--57.

\bibitem[{Woo et~al.(2016)Woo, Kim et~al.}]{woo2016vector}
Woo, J., Kim, N., et~al. (2016).
\newblock Vector field based guidance method for docking of an unmanned surface
  vehicle.
\newblock In \emph{The Twelfth ISOPE Pacific/Asia Offshore Mechanics
  Symposium}. International Society of Offshore and Polar Engineers.

\end{thebibliography}

\appendix
\section{Vessel model} \label{apx:vessel model}
The vessel model used in the simulations was based on the SV Northern Clipper \cite{fossen1996identification}, where the model parameters were taken form the Marine System Simulator (MSS) Toolbox \cite{fossenperez2004}. The model used has the following vessel dynamics
\begin{align*}
    &\dot{\boldsymbol{\eta}} = \boldsymbol{J}(\psi)\boldsymbol{\nu}, \\
    \boldsymbol{M}&\dot{\boldsymbol{\nu}} + \boldsymbol{D}\boldsymbol{\nu} = \boldsymbol{T}(\alpha)\boldsymbol{f}
\end{align*}

%
With the diagonal normalization matrix $\boldsymbol{N}$ = diag([1, 1, L]), and the non-dimensional (bis-system) given by $\boldsymbol{M}_{bis}$ and $\boldsymbol{D}_{bis}$, the mass and dampening matrix are given by the following.
\begin{equation*}
    \boldsymbol{M} = m \boldsymbol{N} \boldsymbol{M}_{bis} \boldsymbol{N}, \quad  \boldsymbol{D} = m \sqrt{\frac{g}{L}} \boldsymbol{N} \boldsymbol{D}_{bis} \boldsymbol{N}
\end{equation*}
\begin{equation*}
    \resizebox{0.9\linewidth}{!}{$
    \boldsymbol{M}_{bis} = 
    \begin{bmatrix}
        1.1274 &        0 &         0 \\
             0 &   1.8902 &   -0.0744 \\
             0 &  -0.0744 &    0.1278
    \end{bmatrix}
    ,\quad
    \boldsymbol{D}_{bis} = 
    \begin{bmatrix}
        0.0358 &       0      &  0      \\
             0 &       0.1183 & -0.0124 \\
             0 &      -0.0041 &  0.0308
     \end{bmatrix}
    $}
\end{equation*}
Where the normalization parameters of length gravity and mass are given as $L = 76.2 (m)$, $g = 9.8 (m/s^2)$ and $m = 6000e3 (kg)$ respectively.

For the vessel, we assume two azimuth thrusters in the aft, with one tunnel thruster in the front giving the thruster position and angle given in Table \ref{tab:thruster_variabels}, and the thrust configuration matrix $\boldsymbol{T}(\boldsymbol{\alpha})$ is as follows.
\begin{equation*}
    \resizebox{.7\linewidth}{!}{$
    \begin{bmatrix}
        \cos(\alpha_1) & \cos(\alpha_2) & 0 \\ 
        \sin(\alpha_1) & \sin(\alpha_2) & 1 \\
        l_{x_1} \sin(\alpha_1) - l_{y_1} \cos(\alpha_1) & l_{x_2} \sin(\alpha_2) - l_{y_2} \cos(\alpha_2) & l_{x_3}
    \end{bmatrix}
    $}
\end{equation*}

\begin{table}[htb!]
    \centering
    \caption{Thruster position and angle}
    \begin{tabular}{|c|c|c|c|}
         \hline
         Truster & $x$-position & $y$-position & angle \\ \hline \hline 
         Azimuth $1$ & $l_{x_1} = - 35m$ & $l_{y_1} = 7m$ & $\alpha_1 $ \\ \hline
         Azimuth $2$ & $l_{x_2} = - 35m$ & $l_{y_2} = -7m$ & $\alpha_2 $ \\ \hline
         Tunnel $3$ & $l_{x_3} = 35m$ & $l_{y_3} = 0m$ & $\alpha_3 = \frac{\pi}{2}$ \\ \hline
    \end{tabular}
    \label{tab:thruster_variabels}
\end{table}

\end{document}